\documentclass[aps,twocolumn,showpacs]{revtex4}
\usepackage{graphics}
\begin{document}
\tighten
\bibliographystyle{apsrev}
\def\half{{1\over 2}}
\def \D {\mbox{D}}
\def\curl {\mbox{curl}\,}
\def \ep {\varepsilon}
\def \lleq {\lower0.9ex\hbox{ $\buildrel < \over \sim$} ~}
\def \ggeq {\lower0.9ex\hbox{ $\buildrel > \over \sim$} ~}
\def\beq{\begin{equation}}
\def\eeq{\end{equation}}
\def\ber{\begin{eqnarray}}
\def\eer{\end{eqnarray}}
\def \apl {ApJ, }
\def \aps {ApJS, }
\def \pd {Phys. Rev. D, }
\def \prl {Phys. Rev. Lett., }
\def \pl {Phys. Lett., }
\def \np {Nucl. Phys., }
\def \l {\Lambda}

\title{A Note on  Inflation with Tachyon Rolling on the Gauss-Bonnet Brane}
\author{B.C. Paul}
\affiliation{ Physics Department, North Bengal University,
 Dist. : Darjeeling, Pin : 734 430, INDIA }
\email{bcpaul@iucaa.ernet.in} 
\author{M.Sami}
\affiliation{ Post Bag 4, Ganeshkhind,\\
 Pune 411 007, India.} 
\altaffiliation[On leave from:]{ Department of Physics, Jamia Millia, New Delhi-110025}
\email{sami@iucaa.ernet.in}
\pacs{98.80.Cq,~98.80.Hw,~04.50.+h}

\begin{abstract}
In this paper we study the tachyonic inflation in brane world cosmology with Gauss-Bonnet term in the bulk. We obtain
the exact solution of slow roll equations in case of exponential potential. We attempt to implement
the proposal of Lidsey and Nunes\cite{g} for the tachyon condensate rolling on the  Gauss-Bonnet brane and discuss the difficulties
associated with the proposal.
\end{abstract}

\maketitle

\section{Introduction}
At present there seems to be no viable alternative
to inflationary scenario. But inspite of all the attractive features of cosmological inflation, 
its mechanism of realization still remains to be ad hoc. As inflation operates at Planck's scale, the needle of hope
points towards the string theory. It is, therefore, not surprising that M/String theory inspired models are under active consideration in cosmology
at present. It was recently been suggested that rolling tachyon condensate, in
a class of string theories, may have interesting cosmological consequences \cite{ashoke}.
Rolling tachyon
matter associated with unstable D-branes has an interesting equation of state
which smoothly interpolates between -1 and 0. 
The tachyonic matter, therefore, might provide an
explanation for inflation at the early epochs and could contribute to some
new form of cosmological dark matter at late time\cite{tachyonindustry}. Unfortunately, this scenario faces difficulties associated
with generation of enough inflation, reheating and the formation of caustics/kinks.\par
   Another interesting development in cosmology inspired by
String theory is related to Brane World cosmology.
In this picture all the matter fields are
confined to the brane whereas gravity can propagate in the
bulk. The scenario has interesting cosmological implications,
in particular, the prospects of inflation are enhanced on the
brane due to the modifications in the
Friedmann equation. While discussing the applications of brane-worlds, one often assumes Einstein
Gravity in the bulk and then projects the dynamics on to the brane. This leads to the high energy
corrections in the Friedman equation which changes the expansion dynamics
in the early universe. To be in the better spirit with string theory, one should include the higher
order curvature invariants to the Einstein-Hilbert action\cite{g1}. The Gauss-Bonnet gravity projected on the brane
leads to modified Friedmann equation different from its counter part in the RS scenario\cite{g2}. And this may
have interesting cosmological consequences. As recently demonstrated by Lidsey and Nunes\cite{g}, the Gauss-Bonnet
modified expansion dynamics can lead to spectral index of perturbation spectrum consistent with the
recent WMAP observation. Interestingly, this is achieved by suitably fixing the Gauss-Bonnet coupling
parameter and the brane tension without tuning the slope of the scalar field potential that drives
inflation. As mentioned above, there are problems with tachyonic inflation as there is no free parameter in the tachyonic potential 
to tune to make the field roll slow allowing the required number of inflationary e-foldings.
 The proposal
of Lidsey and Nunes\cite{g} is specially interesting in case of tachyonic inflation as it does not require the
tuning of the slope of potential.
In this note  we study the tachyonic inflation with exponential potential on Gauss-Bonnet brane and show that 
the spectral index of scalar density perturbations
 can be pushed close to unity. However, for the physical relevant values of parameters, one requires to tune the
slope of the potential.

\subsection{Brane World with Gauss-Bonnet Term in the Bulk}

We consider  five dimensional bulk action with  Gauss-Bonnet term given by  
\begin{eqnarray}
S_{\it M}& =&\nonumber  \frac{1}{2 \kappa_{5}^{2}}\int_M { d^5 x \sqrt{- g} \Big[ R - 2 \Lambda + \alpha ( R^2 - 4 R_{\alpha \beta}  R^{\alpha \beta}} \\          &+&  R_{\alpha \beta \gamma \delta}  R^{\alpha \beta \gamma \delta} ) \Big] + S_{\partial {\it M}} + S_{matter},
\end{eqnarray}
where $\alpha $ represents the Gauss-Bonnet coupling,  $\Lambda $ is the bulk cosmological 
constant. The additive pieces of the action $S_{\partial {\it M}} $  and $S_{matter}$ represent  the action on the boundary and the matter part respectively. The effective 
Friedmann equation is obtained  by imposing a ${\bf Z}_{2}$ symmetry across the brane \cite{g2,g} 
\begin{equation}
H^2 = \frac{c_{+} + c_{-} - 2}{8 \alpha}
\end{equation}
where
\begin{equation}
c_{\pm} = \left( \left[ \left(1 + \frac{4}{3} \alpha \Lambda \right)^{3/2} + \frac{\alpha}{2} \kappa_{5}^{4} \sigma^{2} \right]^{1/2} \pm \sqrt{\frac{\alpha}{2}} \kappa_{5}^{2} \sigma \right)^{2/3},
\end{equation}
and $\sigma$ represents the energy density of the matter sources.

The conservation of energy-momentum on the brane for perfect fluid matter sources is given by
\begin{equation}
\dot{\sigma} + 3 H ( \sigma + p) = 0
\end{equation}
where $p$ represents the pressure of the fluid and $\sigma$ its energy density.

The cosmic dynamics on the brane is fully determined using equations (2) and (4) when the equation of state is known. We use here a simplified approach following Lidsey and Nunes\cite{g}. Accordingly we define a new variable and write the energy density as
\begin{equation}
\sigma = \sqrt{ \left( \frac{2 b}{\alpha \kappa_{5}^{4}} \right)} \; \sinh x,~~~~~~b = \left(1 + \frac{4}{3} \alpha \Lambda \right)^{3/2}      
\label{sigma} 
\end{equation}
Consequently the Friedmann equation (2) takes a simple form which is given by
\begin{equation}
H^{2} = \frac{1}{4 \alpha} \left[ b^{\frac{1}{3}} \cosh \left( \frac{2 x}{3} \right) - 1 \right].
\label{gbfriedman}
\end{equation}
Taking into account (\ref{sigma}) and retaining the terms up to second order in $\sigma$ in 
equation (\ref{gbfriedman}) we have\cite{g}
\begin{equation}
H^{2} = \frac{\kappa_{4}^{2}}{3} \rho  \left[ 1 +  \frac{\rho}{2 \lambda} \right] + \frac{\Lambda_{4}}{3},
\end{equation}
where $\sigma$ is assumed to be decomposed into matter contribution $\rho$
and the brane tension $\lambda$, $\sigma=\lambda+\rho$
and the four-dimensional cosmological constant is given by
\begin{equation}
\Lambda_{4} = \frac{3}{4 \alpha} \left( b^{\frac{1}{3}} - 1 \right) +  \frac{\kappa_{5}^{4}}{12 b^{2/3}} \lambda^{2}.
\end{equation}
The above equation reduces to the standard form of the Friedmann equation  at a sufficiently low energy scales ($\rho << \lambda$) with
\begin{equation}
\kappa_{4}^{2} = \frac{1}{M_{P}^{2}} = \frac{\kappa_{5}^{4} \lambda}{6 b^{\frac{2}{3}}}
\label{kappa}
\end{equation}
where $M_{P}$ is the four-dimensional reduced Planck mass. It is also found that the four dimensional cosmological constant vanishes when the brane tension satisfies
\begin{equation}
\lambda = \frac{3}{2 \alpha \kappa_{4}^{2}}  \left[ 1 - b^{\frac{1}{3}} \right]. \end{equation}

\section{Tachyonic inflation }
The energy density $\rho$
and pressure p for tachyonic field are given by
\begin{equation}
\rho={V(\phi) \over {\sqrt{1-\dot{\phi}^2}}},~~~~~~p=-V(\phi)\sqrt{1-\dot{\phi}^2}
\end{equation}
The field evolution equation (equivalent to  equation (4) is 
\begin{equation}
{\ddot{\phi} \over {1-\dot{\phi}^2}}+3H \dot{\phi}+{V_{,\phi} \over V({\phi})}=0
\label{phieq}
\end{equation}

We now describe inflation on the brane assuming slow roll approximation. Using equations (\ref{sigma}) and (\ref{kappa}) we obtain
\begin{equation}
V =V_0 \sinh x  \equiv \sqrt{ \left( \frac{\lambda  b^{1/3}}{3 \alpha \kappa_{4}^{2}} \right)} \; \sinh x
\label{V} 
\end{equation}
The slow-roll parameters in this case become(we here use the same definition of slow roll
as in Ref\cite{g}.
 
\begin{equation}
\epsilon =  \left( \frac{2 \lambda}{\kappa_{4}^{2}} \frac{V_{,\phi}^2}{V^4} \right) \epsilon_{GB},~~~~
\eta =  \left( \frac{2 \lambda}{\kappa_{4}^{2}}  \left(  \ln V \right),_{\phi \phi} \right) \eta_{GB}       
\end{equation}

where $\epsilon_{GB}$ and $\eta_{GB} $ are given by 
\begin{equation}
\epsilon_{GB} = 
\left[ \frac{2 b^{2/3}}{27} \frac{\sinh (2x/3) \tanh x \sinh^2 x}{(b^{1/3}  \cosh(2x/3) - 1 )^2} \right]
\end{equation}

\begin{equation} 
\eta_{GB} =  \left[ \frac{4 \alpha }{3 (b^{1/3}  \cosh(2x/3) - 1)} \right]
\end{equation}
The number of e-folds of inflationary expansion, is obtained using equations (\ref{gbfriedman}) and 
(\ref{phieq}), which is given by 
\begin{equation}
N(x) = - \frac{3}{4 \alpha} \int_{x_{N}}^{x_{end}} dx \left( \frac{d \phi}{d x} \right)^2 \left( b^{1/3} \cosh (2x/3) - 1 \right) \tanh x
\end{equation}
where $x_{end}$ denotes the value of $x$ when inflation ends. 
The amplitude of scalar perturbation is given by 
\begin{equation}
\delta_{H}^{2} = \left( {1 \over {600 \pi^2}} \frac{\kappa_{4}^{6} V^8}{\lambda^3 V_0 V_{,\phi}^{2}} \right) \left( \frac{729}{8 b} \right) \left( \frac{ (b^{1/3} \cosh (2x/3) - 1 )^3}{\sinh^6 x} \right)
\label{asymmetry}
\end{equation}
using the canonically normalized field  $\phi \rightarrow \sqrt{V_{o}} \phi$ \cite{T}. The exact
expression for the amplitude of density perturbations found by Hwang and Noh\cite{hwang}, $\delta_H={H^2 / {2 \pi \dot{\phi}}} \sqrt{V}$ is not
very different from (\ref{asymmetry} ) as the tachyonic inflation commences very near to the 
top of the potential.

\subsection{Exact Solution of Slow Roll Equations} 
We now study evolution of the universe with tachyon field in an exponential potential which is of the form
\begin{equation}
V = V_{o} e^{-\beta \kappa_{4} \phi},
\end{equation}
It would be interesting to obtain the
exact solutions of slow roll equations in high and low energy limits. In high energy approximation $x>>1$, the Friedmann
equation on the Gauss-Bonnet brane approximates to
\begin{equation}
H^2=A \rho^q
\label{friedman2}
\end{equation}
where $q=2/3$. We shall integrate the equations of motion in slow roll regime for arbitrary $q$. In case of the exponential potential, we have
\begin{equation}
\dot{\phi}(t)={\beta \over{3 \delta M_p}}e^{{{\beta } \over {2M_p}}q\phi}
\label{phidot}
\end{equation}
where $\delta=(A V_0^q)^{1/2}$
Equations (\ref{friedman2}) and (\ref{phidot}) are readily solved to yield
\begin{equation}
\phi(t)=-{{2 M_p} \over {\beta q}}\ln\left[C-{{\beta^2q} \over{6 \delta M_p^2}}t \right]
\label{field}
\end{equation}
\begin{equation}
{a(t) \over a_i}=e^{\delta \left(C-{{\beta^2 q } \over{12 \delta M_p^2}}t \right)t}
\label{sfactor}
\end{equation}

where C is a constant determined by the initial value of $\phi$. The scale factor given by equation (\ref{sfactor}) passes
through a point of inflection marking the end of inflation leading to
\begin{equation}
\dot{\phi}_{end}=\sqrt{{2 \over {3 q}}},~~~~~V_{end}=V_0\left({{\beta q^{1/2}} \over {\sqrt{6} \delta M_p}} \right)^{2/q}
\label{end}
\end{equation}
One can arrive at (\ref{end}) by demanding $\dot{H}+H^2 > 0$ without making use of slow roll equations. Since $ q=2/3$ in the
case under consideration, we find that $\dot{\phi}_{end}=1$. One should
, however keep in mind , that Gauss-Bonnet Friedmann equation in the low energy limit ($x<<1$ ) reduces to standard
RS form and in case $\rho>>\lambda$ reduces to (\ref{friedman2}) with $q=2$ and $\delta=(1/6\lambda M_p^2)^{1/2}$ leading to the following\cite{T}
\begin{equation}
\dot{\phi}_{end}=\sqrt{{1 \over 3}},~~~~~V_{end}=\beta \sqrt{{2 \lambda}}
\label{infend}
\end{equation}
 
\subsection{Number of e-foldings and Density Perturbations}

We now compute the number of e-folds and density perturbation for the tachyonic system.
For an exponential potential, the value of the field, $x_{N}$, corresponding to $N$ e-folds before the end of inflation is given by 
\begin{equation}
N = - \frac{3}{4 \alpha} \int_{x_{N}}^{x_{end}} dx \frac{1}{(\beta \kappa_{4})^2} \coth^2 x \left( b^{1/3} \cosh (2 x /3) - 1 \right) \tanh x.
\label{integral}
\end{equation}
The integral in (\ref{integral}) may be evaluated analytically, which becomes
\begin{equation}
N = - \frac{3}{4 \alpha (\beta \kappa_{4})^2} \left[ f(x) \right]_{x_{N}}^{x_{end}}
\label{N}
\end{equation}
where we have
\begin{eqnarray}
f(x)& =& \nonumber \frac{3}{2} b^{1/3} \cosh (2x/3) - \\
    &-& \nonumber \frac{1}{2} (2 + b^{1/3}) \ln [1 + 2 \cosh (2 x/3)] +\\ 
    &+& (b - 1) \ln( \sinh (x/3))
\label{f} 
\end{eqnarray}
It is possible to simplify the above expression if $\alpha \Lambda << 1$ which leads to $b \sim 1$. The  end of the inflationary epoch  is determined by noting that $x >> 1$, the slow roll parameter $\epsilon >>1$. Thus  one determines $x_{end}$ and $V_{end}$ from the condition that $x$ is sufficiently small (i.e., $x << 1$). Using equations (\ref{V}) and (\ref{infend}), it is now possible to determine $x_{end}$ 
\begin{equation}
x_{end} = (6 \alpha \beta^2 \kappa_{4}^2)^{1/2}.
\label{xend}
\end{equation}
Using equation (\ref{f}) gives
\begin{equation}
f(x_{end})={3 \over 2}-{{3 \ln(3)} \over 3}+{x_{end}^2 \over 9}
\label{f(xend)}
\end{equation}
Now equating  (\ref{xend}) and (\ref{N}) we determine $\alpha$ 
\begin{equation}
\alpha = \frac{3}{4 (2 N + 1) \beta^2 \kappa_{4}^{2}} \left[ 3 (\ln 3 - 1) + 2 f(x_{N}) \right]
\label{alpha1}
\end{equation}
Using (\ref{asymmetry}) and COBE normalization we obtain
\begin{equation}
\alpha^5 \lambda = \frac{243 \times 10^{16}}{4096 \kappa_{4}^{2} \pi^2} \frac{1}{\beta^4} \left[ \cosh (2 x_{N}/3) - 1 \right]^3
\label{alpha2} 
\end{equation}
One may now extract the values of the brane tension $\lambda$ and Gauss-Bonnet coupling, $\alpha$ which are consistent with the COBE normalization.\\ 

The slow roll parameters can now be cast in the form
\begin{equation}
\epsilon = \left[ \frac{\ln 3 - 1 + \frac{2}{3} f(x_{N})}{2 N + 1} \right] \frac{\sinh (2 x_{N}/3) \tanh (x_{N})}{(\cosh (2 x_{N}/3) - 1)^2}
\label{epcilonf}
\end{equation}
while $\eta=0$ in case of exponential potential. To get scalar field confined on the brane we have to impose the constraint $\rho < \kappa_{5}^{-8/3}$ which leads to a lower limit on  the allowed values of $\alpha^3 \lambda $ for a given $x_{N}$\cite{g}
\begin{equation}
\alpha^3 \lambda  > 48 \kappa_{4}^{2} \sinh^6 x_{N}.
\label{ineq}
\end{equation}
Using equations (\ref{alpha1}) and (\ref{alpha2}) we determine
\begin{eqnarray}
\alpha^3 \lambda& =& \frac{27 \times 10^{16}}{256 \pi^2} \kappa_{4}^{2}\left(\frac{2 N + 1}{ 3(ln 3 - 1) + 2 f(x_{N}) } \right)^2\\ 
 &\times& \nonumber \left( \cosh (2 x_{N}/3) - 1 \right)^6
\end{eqnarray}
The inequality (\ref{ineq}) is satisfied for values of $x_N$ such that $x_N \le 6.5$. Making use of (\ref{epcilonf}), we find that 
the spectral index $n_s \simeq 0.97$ for an allowed value of
 $x_N$. Unfortunately, the numerical values of $\lambda$ and $V_{end}$ turn out to be larger than the Planck's scale 
if $\beta/M_p^2 \sim 1$ , i.e., if one restricts to string theory tachyons. This ,of course, puts the whole
procedure under doubt as the classical treatment of gravity becomes invalid in this case. 
Actually, the problem is present even at the RS level itself which, it seems, can not be corrected by 
Gauss-Bonnet correction. Indeed, the RS level expression for the asymmetry is given by,
\begin{equation}
\delta_H^2={{8 k_4^4 \beta^4 \times (2N+1)^{5/2}} \over {600 \pi^2 V_{end}}} \simeq 10^{-10}
\label{AsRS}
\end{equation}
where $V_{end} =V/(2N+1)^{1/2}$\cite{T} in case of the exponential potential. It follows from equation (\ref{AsRS}) that $V_{end} > M_p^4$ if $\beta/M_p^2 \simeq 1$.This is related to the fact that not enough inflation can be drawn in case
of tachyonic field if one restricts to string theory tachyons.
 The situation could be remedied either by tuning the parameter $\beta$ or by introducing the large number
of D-branes parallel to each other and separated by a distance much larger than $l_s$(this is
the same mechanism of assisted inflation as introduced by Liddle et. al\cite{Liddle}). The first option is out of place as $\beta$
is not a free parameter in tachyon potential($\beta$ is fixed by sring scale $l_s$). As for the large number of D-brane assisted inflation is concerned,
one could draw enough inflation in this procedure and push the spectral index $n_s$ close to one in the standard FRW cosmology itself.
The sole purpose of invoking the brane worlds with Gauss-Bonnet term in the bulk is defeated. It is remarkable that the exponential
potential, in case of normal scalar field, on the brane with Gauss-Bonnet Einstein equations in bulk allows to push  $n_s$ very close to one independently of the slope
of potential. It was, therefore, natural to investigate the tachyonic system along the same line as there is no free parameter in this case to tune.
But due to the peculiar nature of tachyon field dynamics, the proposal of Lidsey and Nunes does not seem to work  in a natural way.
 
\section*{Acknowledgments} We are thankful to J. E. Lidsey, N. J. Nunes and N. Dadhich for their critical comments. We also thank James Gregory, S. Odintsov  and S. Panda for helpful comments.


\begin{thebibliography}{99}
\bibitem{g}J. E. Lidsey and N. J. Nunes, {\it Phys. Rev. } {\bf D 67}, 103510 (2003).
\bibitem{ashoke} A. Sen, hep-th/0203211; hep-th/0203265; hep-th/0204143.
\bibitem{tachyonindustry}
Gibbons, G.W, arXiv:hep-th/0204008;  
A. Mazumdar, S. Panda, A. Perez-Lorenzana, hep-ph/0107058; 
Mukohyama, S hep-th/0204084;
Frolov, A, L.~Kofman and A.~Starobinsky, Phys.Lett. B{\bf 545}, 8 (2002);
Choudhury, D, D.~Ghoshal, D.~P.~Jatkar and S.~Panda, hep-th/0204204;
    G Shiu and I. Wasserman,
    Phys.Lett. B541 (2002) 6;
       Padmanabhan, T., Phys. Rev. D{\bf 66}, 021301(2002); T. Padmanabhan, T. Roy Choudhury, hep-th/0205055; 
        Kofman, L and A. Linde, arXiv: hep-th/0205121;
	 Sami, M., hep-th/0205146;
	      Piao, Y.S, R.~G.~Cai, X.~m.~Zhang and Y.~Z.~Zhang, hep-ph/0207143;
	        Cline, J.M, H.~Firouzjahi and P.~Martineau, hep-th/0207156;
		  Zong-Kuan Guo, Yun-Song Piao, Rong-Gen Cai, Yuan-Zhong Zhang,  hep-ph/0304236; 
		  Yun-Song Piao, Rong-Gen Cai, Xinmin Zhang, Yuan-Zhong Zhang, hep-ph/0207143;
		 G. Felder, Lev Kofman and A. Starobinsky, JHEP 0209 (2002) 026.
		 Wang, B, E.~Abdalla and R.~K.~Su, hep-th/0208023;
		     Jian-gang Hao, Xin-zhou Li,  hep-th/0209041;  
		       M. C. Bento, O. Bertolami and A.A. Sen, hep-th/020812;  M.C. Bento, O. Bertolami., 
		       A.A. Sen, Phys.Rev.D67:023504,2003;
		           gr-qc/0204046; M.C. Bento, O. Bertolami., Phys.Rev.D65:063513,2002;  
			   Jian-gang Hao, Xin-zhou Li, Phys.Rev. D66 (2002) 087301;
			   Chanju Kim , Hang Bae Kim and Yoonbai Kim,
			   hep-th/0210101;  
			   J.S.Bagla,  H.K.Jassal, T.Padmanabhan, astro-ph/0212198;
			    M. Sami, P. Chingangbam and T. Qureshi, hep-th/0301140;
			     Xin-zhou Li, Dao-jun Liu, Jian-gang Hao, hep-th/0207146;  
			      M. Majumdar, Anne-Christine Davis, hep-th/0304226; 
                              D. Choudhury, D. Ghoshal,  
			      D. P. Jatkar, S. Panda, hep-th/0305104;  
                             D.A.Steer, F.Vernizzi, hep-th/0310139;
			      Zong-Kuan Guo, Hong-Sheng Zhang, Yuan-Zhong Zhang,  hep-ph/0309163 ; P.K.Suresh, gr-qc/0309043;
			      M. R. Garousi,hep-th/0307197; Maria C. Bento, Nuno M. C. Santos, A. A. Sen, astro-ph/0307292;
			       Yun-Song Piao, Yuan-Zhong Zhang,. hep-th/0307074 ; D. Bazeia, 
			       F.A. Brito, J.R. Nascimento, hep-th/0306284;
			        Shin'ichi Nojiri, Sergei D. Odintsov, hep-th/0306212; Jian-gang Hao, Xin-zhou Li,. hep-th/0305207;
				L. Raul W. Abramo, Fabio Finelli, astro-ph/0307208; V. Gorini, A. Yu. Kamenshchik and U. Moschella, V. Pasquier, hep-th/0311111; Marie-Bernadette Causse, astro-ph/0312206;
Dao-jun Liu, Xin-zhou Li,
astro-ph/0402063; J.M. Aguirregabiria, Ruth Lazkoz,  hep-th/0402190;  J.M. Aguirregabiria, Ruth Lazkoz, gr-qc/0402060 ;  Gianluca Calcagni, hep-ph/0402126; Kenji Hotta, hep-th/0403078 ; Xin-He Meng, Peng
Wang, hep-ph/0312113; M. R. Garousi, M. Sami and Shinji Tsujikawa, hep-th/0402075; Luis P. Chimento,
astro-ph/0311613.
 
\bibitem{T}Fairbairn M and M.~H.~Tytgat, hep-th/0204070;  Sami, M, P.~Chingangbam and T.~Qureshi, hep-th/0205179.
\bibitem{g1}
B. Zwiebach, in {\it Anomalies, Geometry and Topology} - proceedings of the symposium, Argonne, Illinois, 1985, edited by W. A. Bar; 
D. Lorentz-Petzold, {\it Prog. Theor. Phys. } {\bf 78}, 969 (1987);{\it Mod. Phys. Lett.} {\bf A 3}, 827  (1988);
 A. B. Henriques, {\it Nucl. Phys. } {\bf 277 B}, 621 (1986);
Q. Shafi and C. Wetterich, {\it Phys. Lett.} {\bf 129 B}, 387 (1983);  {\it ibid} {\bf 152 B}, 51 (1984);
H. Ishihara, {\it Phys. Lett.} {\bf 179 B}, 217 (1986);
B. C. Paul and S. Mukherjee, {\it Phys. Rev.} {\bf D 42}, 2595 (1990);
I. P. Neupane, hep-th/0108194; I. P. Neupane,  hep-th/0106100.
\bibitem{g2}Shin'ichi Nojiri, Sergei D. Odintsov, hep-th/0006232; B. Abdesselam and N. Mohammedi, {\it Phys. Rev.} {\bf D 65}, 084018 (2002); J. E. Lidsey, Shin'ichi Nojiri, S. D. Odintsov, hep-th/0202198; 
C. Charmouses and J. Dufaux, hep-th/0202107; S.C. Devis, hep-th/0208205; E. Gravanis and S. Willison, hep-th/0209076; S. Nojiri, S.Odintsov and S. Ogushi, Phys. Rev. D {\bf 65}
(2002) 023521; J. P. Gregory, A. Padilla, hep-th/0304250;  
H. M. Lee, hep-th/0010193 (2000); J. E. Kim and H. M. Lee, gr-qc/0306116 (2003);
     N. Deruelle and C. Germani, gr-qc/0306116 (2003); Nathalie Deruelle and John Madore,  gr-qc/0305004;  
      Kei-ichi Maeda, Takashi Torii, hep-th/0309152. 
\bibitem{hwang}  J. Hwang and H. Noh,  hep-th/0206100.
\bibitem{Liddle} Andrew R. Liddle, A. Mazumdar, F. E. Schunck, astro-ph/9804177.
\end{thebibliography}
\end{document}